\documentclass{article}
\usepackage{amssymb}
\usepackage{axodraw}
\usepackage{latexsym}
\usepackage{epsfig}

\title{\large \bf Field Decomposition and the Ground State Structure of SU(2) Yang-Mills Theory}
\author{\vspace{-0.1cm}\normalsize Lisa Freyhult\footnote{lisa.freyhult@teorfys.uu.se}\\ \vspace{-0.1cm}\normalsize \textsl{Department
of Theoretical Physics, Uppsala University}\\\normalsize \textsl{P.O. Box 803, S-75108, 
Uppsala, Sweden}} 

\date{}

\begin{document}
\maketitle

\begin{abstract}
\noindent
We compute the effective potential of SU(2) Yang-Mills theory using the background field
method and the Faddeev-Niemi decomposition of the gauge fields. 
In particular, we find that the potential will depend on the values of two scalar fields in the decomposition  
and that its structure will give rise to a symmetry breaking.
\end{abstract} 
Recently it has been proposed that the different phases of Yang-Mills
theory can be described by an appropriate decomposition of the gauge
fields \cite{Antti2}. This
decomposition has been used to describe the long distance limit
of Yang-Mills theory, where stable knotted
solitons have been found \cite{Antti3}. Furthermore the
decomposition leads to a Lagrangian with a manifest duality between the
electric and magnetic variables \cite{Antti}. 

Here we
shall investigate the effective theory in terms of the variables of the
decomposition. The result will be an alternative way to view
the ground state structure of the theory. We find that the
effective potential, at the one loop level of approximation, will lead to
a non-trivial ground state which suggests that symmetry breaking and
dimensional transmutation \cite{Coleman-Weinberg} takes place.

In order to determine the effective theory we rely on the formalism of
Coleman and Weinberg \cite{Coleman-Weinberg} and we will
use the background field method \cite{Abbott} to perform the
calculations. 

The classical action of SU(2) Yang-Mills theory is
\begin{equation}
S=-\frac{1}{4}\int d^4xF^a_{\mu\nu}F^{a\mu\nu}
\end{equation}
where $F_{\mu\nu}^a=\partial_\mu A_\nu^a-\partial_\nu
A_\mu^a+g\epsilon^{abc}A_\mu^bA_\nu^c$. 
Following \cite{Antti2}\cite{Antti} the off diagonal gauge fields are decomposed as follows.
\begin{equation}
A_\mu^+=A_\mu^1+iA_\mu^2=i\psi_1\mathbf{e}_\mu+i\psi_2\mathbf{e}_\mu^*
\end{equation}
where
\begin{equation}
\mathbf{e}_\mu=\frac{1}{\sqrt2}(e_\mu^1+ie_\mu^2)
\end{equation}
with the normalisation condition
\begin{equation}
e_\mu^ae^{b\mu }=\delta^{ab}
\end{equation}
$\psi_1$ and $\psi_2$ are complex scalar fields. The diagonal gauge field, $A_\mu^3$, remain intact. 
Here we are interested in implementing this decomposition in the one-loop
effective Yang-Mills action \cite{Drummond}\cite{Savvidy}. For this we first use the
background
field method \cite{Abbott} to find the effective potential up to one-loop order and then introduce the change of variables
as in (2). This will yield an effective potential dependent only on the scalar
fields in the decomposition, which allows us to study the phase structure
of the theory.

In order to implement the background field method we make the following shift of the gauge fields in the action
\begin{equation}
A_\mu^a\rightarrow A_\mu^a+\mathcal{A}_\mu^a
\end{equation}
where we view $A_\mu^a$ as a classical background field. When quantising, using the path integral formulation, we integrate
over $\mathcal{A}_\mu^a$ only. 
We shall find that 
making this shift corresponds to making the following shifts in the complex scalar fields
\begin{equation}
\psi_1\rightarrow f_0+\psi_1\quad \psi_2\rightarrow g_0+\psi_2\quad f_0,g_0\in\mathbb{R}
\end{equation}
where $f_0$ and $g_0$ are constants. This is due to the fact that the minima of the eventual potential should be
translation invariant. Hence we conclude that the shifted off diagonal fields are
\begin{equation}
A_\mu^1=\frac{1}{\sqrt2}(g_0-f_0)e_\mu^2\quad A_\mu^2=\frac{1}{\sqrt2}(f_0+g_0)e_\mu^1
\end{equation}

The effective potential is the negative sum of all non-derivative
terms in the Lagrangian. We remove the derivatives in a covariant way,
i.e. we also remove the background field $A_\mu^3$. At the tree level of
Yang-Mills we have
\begin{equation}
V=\frac{g^2}{4}\epsilon^{abc}\epsilon^{aef} A_\mu^bA_\nu^cA^{\mu e}A^{\nu f}
\end{equation}
This quantity is in general not gauge invariant. However it is invariant under constant gauge transformations,
which is consistent with our expectation that the ground state of the
theory should be translation invariant.
Using the decomposition (7) we obtain
\begin{equation}
V=\frac{g^2}{8}(f_0^2-g_0^2)^2
\end{equation}
The minima of this potential is along the directions $f_0=\pm g_0$ where
the potential is zero. The vacuum is infinitely degenerate and there is
no symmetry breaking. In the following we are interested in locating the
ground state of the theory by inspecting radiative corrections to (9). 

 The Lagrangian
is invariant under gauge transformations of $\mathcal{A}_\mu^a$ and in
order to compute the quantum corrections we fix the
gauge. 
Here we will use the background field analogue of Feynman gauge, the Lagrangian with gauge fixing terms and ghosts is then
\begin{eqnarray}
L\nonumber&=&-\frac{1}{4}\left(\partial_\mu(A_\nu^a+\mathcal{A}_\nu^a)-\partial_\nu(A_\mu^a+\mathcal{A}_\mu^a)+g\epsilon^{abc}(A_\mu^b+\mathcal{A}_\mu^b)(A_\nu^c+\mathcal{A}_\nu^c)\right)^2
\\&&-\frac{1}{2\xi }\left((\partial_\mu\delta^{ab}+g\epsilon^{abc}A^c_\mu)\mathcal{A}_\mu^a
\right)^2-
\bar{c}^a(\partial_\mu\delta^{ab}+g\epsilon^{abc}(A_\mu^c+\mathcal{A}_\mu^c))^2c^c
\end{eqnarray}
Where $\xi$ is the gauge fixing parameter, here $\xi=1$.
The Lagrangian can be rewritten, keeping only terms quadratic in $\mathcal{A}_\mu^a$ and $c^a$, as follows
\begin{eqnarray}
L&=&-\frac{1}{2}\mathcal{A}^{a\alpha }\left(-(D^2)^{ac}g_{\alpha\beta}-
2g\epsilon^{abc}F^b_{\alpha\beta}\right)\mathcal{A}^{c\beta}+\bar{c}^a(-
(D^2)^{ac})c^c
\end{eqnarray}
where $D_\mu^{ab}=\partial_\mu\delta^{ab}+g\epsilon^{abc}A_\mu^c$. We
note here the gauge invariant structure of the differential operators
with respect to background gauge transformations. The
motivation for keeping quadratic terms only is that the linear terms
disappear in the calculation process 
and higher
powers than two in $\mathcal{A}_\mu^a$ do not contribute to the one-loop
corrections. 

The effective action, $S_{eff}$, in Euclidean space, is defined by 
\begin{eqnarray}
\nonumber e^{S_{eff}}&=&\int [d\mathcal{A}][dc][d\bar{c}]\exp\left(\int d^4x L\right)\\
&=&\exp\left(\int d^4x(-\frac{1}{4}(F^a_{\mu\nu})^2+...)\right)(\mbox{det}\Delta_{gauge})^{-1/2}(\mbox{det}\Delta_{ghost})
\end{eqnarray}

We note that the result of the computation will be a renormalisation
of the gauge coupling and the background field.
\begin{equation}
g\rightarrow Z_gg\quad A_\mu\rightarrow Z_A^{1/2}A_\mu
\end{equation}
Also the gauge fixing
parameter, $\xi$, is renormalised but that we need not consider when computing one-loop corrections.
Gauge invariance dictates  that $Z_g=Z_A^{-1/2}$ \cite{Abbott} so that the effective Lagrangian will look like
\begin{equation}
L_{eff}=-\frac{1}{4}Z_AF_{\mu\nu}^aF^{a\mu\nu}+\mbox{higher order terms}
\end{equation} 
To compute the
functional determinants in (12) the proper time method \cite{Schwinger}
is often used. See for example \cite{Nielsen-Olesen}, \cite{Savvidy}
and \cite{Schanbacher}. Here we will however use a different method,
similar to \cite{Coleman-Weinberg} and \cite{Jackiw}. We feel this
method might simplify our computations here.

Computing the functional determinants in equation (12) corresponds
to computing the Feynman diagrams in figure 1 and the corresponding ghost diagrams.

\begin{figure}[h]
\vspace{-1cm}
\begin{picture}(50,50)
\Gluon(0,0)(20,0){2}{4}
\Vertex(20,0){1}
\GlueArc(30,0)(10,0,180){2}{6}
\Vertex(40,0){1}
\GlueArc(30,0)(10,180,360){2}{6}
\Gluon(40,0)(60,0){2}{4}
\PText(70,0)(0)[c]{+}
\Gluon(80,0)(100,0){2}{4}
\Vertex(100,0){1}
\GlueArc(110,0)(10,0,90){2}{3}
\Gluon(110,10)(110,30){2}{4}
\Vertex(110,10){1}
\GlueArc(110,0)(10,90,180){2}{3}
\Vertex(120,0){1}
\GlueArc(110,0)(10,180,270){2}{3}
\Vertex(110,-10){1}
\Gluon(110,-10)(110,-30){2}{4}
\GlueArc(110,0)(10,270,360){2}{3}
\Gluon(120,0)(140,0){2}{4}
\PText(145,0)(0)[cl]{+. . .+}
\GlueArc(190,0)(10,-90,270){2}{13}
\Vertex(190,-10){1}
\Gluon(180,-20)(190,-10){2}{3}
\Gluon(190,-10)(200,-20){2}{3}
\PText(215,0)(0)[c]{+}
\GlueArc(240,0)(10,-90,90){2}{6}
\Vertex(240,-10){1}
\Gluon(230,-20)(240,-10){2}{3}
\Gluon(240,-10)(250,-20){2}{3}
\GlueArc(240,0)(10,90,270){2}{6}
\Vertex(240,10){1}
\Gluon(230,20)(240,10){2}{3}
\Gluon(240,10)(250,20){2}{3}
\PText(270,0)(0)[c]{+. . .}
\PText(0,-40)(0)[l]{+ diagrams that mixes cubic and quartic vetrices}
\end{picture}
\vspace{1.5cm}
\caption{The types of diagrams contributing to the effective action at the one-loop level of approximation.}
\vspace{-0.3cm}
\end{figure}
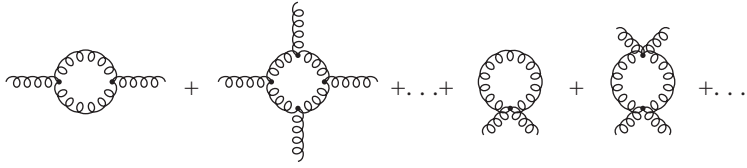
\newpage
\noindent
Starting with the first series of diagrams we obtain
\begin{figure}[h]
\vspace{0.7cm}
\hspace{0.7cm}
\begin{picture}(1,1)
\Gluon(0,0)(20,0){2}{4}
\Vertex(20,0){1}
\GlueArc(30,0)(10,0,180){2}{6}
\Vertex(40,0){1}
\GlueArc(30,0)(10,180,360){2}{6}
\Gluon(40,0)(60,0){2}{4}
\PText(70,0)(0)[c]{+}
\Gluon(80,0)(100,0){2}{4}
\Vertex(100,0){1}
\GlueArc(110,0)(10,0,90){2}{3}
\Gluon(110,10)(110,30){2}{4}
\Vertex(110,10){1}
\GlueArc(110,0)(10,90,180){2}{3}
\Vertex(120,0){1}
\GlueArc(110,0)(10,180,270){2}{3}
\Vertex(110,-10){1}
\Gluon(110,-10)(110,-30){2}{4}
\GlueArc(110,0)(10,270,360){2}{3}
\Gluon(120,0)(140,0){2}{4}
\PText(145,0)(0)[cl]{+. . .}
\end{picture}
\vspace{0.7cm}
\end{figure}
\begin{eqnarray}
\nonumber&=&\mbox{Tr}\int\frac{d^4k}{(2\pi)^4}\int\frac{d^4p}{(2\pi)^4}\sum_{n=
1}^{\infty}\frac{1}{2n}\bigg(\frac{i}{k^2}\Big(2gg_{\alpha\beta}\epsilon^{acd}A_\mu^
dk_\mu+gg_{\alpha\beta}\epsilon^{acd}A_\mu^dp_\mu\\
&&+2g\epsilon^{abc}(p_\alpha
A_\beta^b-p_\beta A_\alpha^b)\Big)\bigg)^n
\end{eqnarray}
Note that the trace is over both the greek and latin indices.
The diagrams with an odd number of vertices disappear when evaluating the trace.
The expression is a geometric series, which we can rewrite in the following way:
\begin{equation}
\sum_{n=1}^{\infty}\frac{x^n}{n}=-\ln(1-x)=-\frac{\partial}{\partial\alpha}\frac{1}{(1-x)^\alpha}\bigg|_{\alpha=0}
\end{equation}
Exchanging the order of the derivative with respect to $\alpha$ and the integration over $k$ (15) take the following form
\begin{eqnarray}
&&\nonumber-\frac{20}{3}\frac{1}{4}\frac{1}{(4\pi)^2}\mbox{Tr}\int\frac{d^4p}{\left(2\pi\right)^4}g^2\left(\delta^{ag}A_\mu^dA_\nu^d-
A_\mu^aA_\nu^g\right)\left(p^2g^{\mu\nu}-
p^\mu p^\nu\right)\\
&&\times\ln\left(\frac{1}{\mu_1^2}\sqrt{g^2\left(\delta^{gc}A_\gamma^eA_\delta^e-
A_\gamma^gA_\delta^c\right)\left(p^2g^{\gamma\delta}-
p^\gamma p^\delta\right)}\right)+\cdots
\end{eqnarray}
We write out corrections to the Abelian part of the action (that is $-\frac{1}{4}\int d^4x(\partial_\mu A^a_\nu-\partial_\nu A^a_\mu)^2$) and we only
keep terms of the order $p^2$ inside the logarithm. The other terms
will be obtained eventually by arguments of gauge invariance, using equation (14) etc. Here $\mu_1$ is a
mass scale.
The second type of diagrams contributing to the effective action is
\begin{figure}[h]
\vspace{0.7cm}
\hspace{0.3cm}
\begin{picture}(1,1)
\GlueArc(10,0)(10,-90,270){2}{13}
\Vertex(10,-10){1}
\Gluon(0,-20)(10,-10){2}{3}
\Gluon(10,-10)(20,-20){2}{3}
\PText(35,0)(0)[c]{+}
\GlueArc(60,0)(10,-90,90){2}{6}
\Vertex(60,-10){1}
\Gluon(50,-20)(60,-10){2}{3}
\Gluon(60,-10)(70,-20){2}{3}
\GlueArc(60,0)(10,90,270){2}{6}
\Vertex(60,10){1}
\Gluon(50,20)(60,10){2}{3}
\Gluon(60,10)(70,20){2}{3}
\PText(90,0)(0)[c]{+. . .}
\end{picture}
\vspace{0.3cm}
\end{figure}

\noindent
These diagrams give no contribution to the Abelian part of the action and
since we rely on the arguments of gauge invariance we need not compute
these type of diagrams. The same is true for the corresponding ghost
diagrams. Furthermore there is no need to compute the diagrams that
mixes cubic and quartic vertices since they do not contribute to the
Abelian part of the effective action.

\begin{figure}[h]
\mbox{We compute the first type of ghost diagrams}
\newline
\vspace{-0.5cm}
\hspace{0.9cm}
\begin{picture}(40,40)
\Gluon(0,0)(20,0){2}{4}
\Vertex(20,0){1}
\DashCArc(30,0)(10,0,180){2}
\Vertex(40,0){1}
\DashCArc(30,0)(10,180,360){2}
\Gluon(40,0)(60,0){2}{4}
\PText(70,0)(0)[c]{+}
\Gluon(80,0)(100,0){2}{4}
\Vertex(100,0){1}
\DashCArc(110,0)(10,0,90){2}
\Gluon(110,10)(110,30){2}{4}
\Vertex(110,10){1}
\DashCArc(110,0)(10,90,180){2}
\Vertex(120,0){1}
\DashCArc(110,0)(10,180,270){2}
\Vertex(110,-10){1}
\Gluon(110,-10)(110,-30){2}{4}
\DashCArc(110,0)(10,270,360){2}
\Gluon(120,0)(140,0){2}{4}
\PText(145,0)(0)[cl]{+. . .}
\end{picture}
\vspace{0.7cm}
\end{figure}

\begin{eqnarray}
&=&-\mbox{Tr}\int\frac{d^4k}{(2\pi)^4}\int\frac{d^4p}{(2\pi)^4}\sum_{n=
1}^{\infty}\frac{1}{n}\bigg(\frac{i}{k^2}\Big(2g\epsilon^{acd}A_\mu^
dk_\mu+g\epsilon^{acd}A_\mu^dp_\mu\Big)\bigg)^n\\
\nonumber&=&-\frac{2}{3}\frac{1}{4}\frac{1}{(4\pi)^2}\mbox{Tr}\int\frac{d^4p}{\left(2\pi\right)^4}g^2\left(\delta^{ag}A_\mu^dA_\nu^d-
A_\mu^aA_\nu^g\right)\left(p^2g^{\mu\nu}-
p^\mu p^\nu\right)\\
&&\times
\ln\left(\frac{1}{\mu_2^2}\sqrt{g^2\left(\delta^{gc}A_\gamma^eA_\delta^e-
A_\gamma^gA_\delta^c\right)\left(p^2g^{\gamma\delta}-
p^\gamma p^\delta\right)}\right)+\cdots
\end{eqnarray}
where $\mu_2$ is a constant of dimension mass. 

Collecting the results from the computation of the diagrams we obtain
\begin{eqnarray}
\nonumber S_{eff}&=&-
\frac{1}{4}\mbox{Tr}\int\frac{d^4p}{(2\pi)^4}(\delta^{ag}A_\mu^dA_\nu^d-A_\mu^aA_\nu^g)(p^2g^{\mu\nu}-p^\mu
p^\nu)\\
&&\nonumber\times\left(\delta^{gc}+\frac{22}{3}\frac{g^2}{(4\pi)^2}\ln\left(\frac{1}{\mu^2}\sqrt{g^2(\delta^{gc}A_\gamma^eA_\delta^e-A_\gamma^gA_\delta^c)(p^2g^{\gamma\delta}-p^\gamma
p^\delta)}\right)\right)\\&&+\cdots
\end{eqnarray}
We complete this into a gauge invariant quantity using (14) and general
arguments of gauge invariance. We conclude that the result should be of
the following form
\begin{equation}
S_{eff}=-
\frac{1}{4}F^a_{\mu\nu}F^{b\mu\nu}\left(\delta^{ab}+\frac{22}{3}\frac{g^2}
{(4\pi)^2}\ln\left(\frac{1}{\mu^2}\sqrt{g^2F^a_{\gamma\delta}F^{b\gamma\delta}}\right)\right
)
\end{equation}
Note that here we have rewritten the expression without the trace for
simplicity. So far we have done the calculations in Euclidean space but
now we make an analytic continuation to Minkowski space to be able to compare our results to those of others \cite{Cho1}, \cite{Cho2} and \cite{Schanbacher}.
In terms of the colour electric and magnetic fields, defined as $E^a_i=F^a_{i0}$ and $B^a_i=\frac{1}{2}\epsilon_{ijk}F^a_{jk}$, the effective action is
\begin{equation}
S_{eff}=\frac{1}{2}(E^aE^b-B^aB^b)\left(\delta^{ab}+\frac{22}{3}\frac{g^2}
{(4\pi)^2}\ln\left(\frac{1}{\mu^2}\sqrt{2g^2(B^aB^b-E^aE^b)}\right)\right
)
\end{equation}
For a purely electric field we have:
\begin{equation}
S_{eff}=\frac{1}{2}E^aE^b\left(\delta^{ab}+\frac{22}{3}\frac{g^2}
{(4\pi)^2}\ln\left(\frac{1}{\mu^2}\sqrt{2g^2E^aE^b)}\right)-\frac{11i}{48\pi}g^2\delta^{ab}\right
)
\end{equation}
However for a purely magnetic field we have
\begin{equation}
S_{eff}=-\frac{1}{2}B^aB^b\left(\delta^{ab}+\frac{22}{3}\frac{g^2}
{(4\pi)^2}\ln\left(\frac{1}{\mu^2}\sqrt{2g^2B^aB^b}\right)\right
)
\end{equation} 
If we consider a background field on the form $F_{\mu\nu}^a(x)=n^aF_{\mu\nu}(x)$ (the background field that Savvidy choose \cite{Savvidy})
where $n^a$ is a constant colour vector (and $n^an^a=1$) we see that there is no imaginary
term in (24) while there is one in (23).
These results have been obtained by several authors
\cite{Cho1}\cite{Cho2}\cite{Schanbacher} using the proper time method \cite{Schwinger}. The instability corresponding to the
imaginary term in the action was first discussed by Nielsen and Olesen
\cite{Nielsen-Olesen}. It was later modified by chosing the integration
path in the proper time integral by arguments of causality
\cite{Cho2}\cite{Schanbacher}. Our result agrees with these later results and provides an independent justification of these arguments.

Here we are interested in the low momentum limit of the effective action in (21) which we identify as the effective potential 
\begin{eqnarray}
\nonumber V_{eff}&=&\frac{g^2}{4}\epsilon^{acd}\epsilon^{bef}A_\mu^cA_\nu^dA^{e\mu}A^{f\nu}\\&&\times\left(\delta^{ab}+\frac{22}{3}\frac{g^2}{(4\pi)^
2}\ln\left(\frac{1}{\mu^2}(\sqrt{g^4\epsilon^{agh}\epsilon^{bij}A_\gamma^gA_\delta^hA^{i\gamma}A^{j\delta}}\right)\right)
\end{eqnarray}
We now want to inspect the properties of this potential using the decomposed variables.
 Making the substitution (7) in (25) gives 
\begin{eqnarray}
V_{eff}&=&\frac{g^2}{8}(f_0^2-
g_0^2)^2+\frac{22}{3}\frac{1}{16\pi^2}\frac{g^4}{8}(f_0^2-
g_0^2)^2\ln\left(\frac{g^2}{\mu^2\sqrt{2}}|f_0^2-g_0^2|\right)
\end{eqnarray}
Notice that the effective potential is a function only of the two scalar
fields in the decomposition and that it is completely symmetric. There
is no imaginary term present here. This is a consequence of the choice
of background field in equations (4) and (7).

In (26) we have the cutoff scale $\mu$ and in order to exchange it for
the renormalisation scale we choose to define the coupling constant as
\begin{equation}
g^2=\frac{1}{3}\frac{\partial^4V_{eff}}{\partial
f_0^4}\bigg|_{(f_0,g_0)=(M,0)}=
\frac{1}{3}\frac{\partial^4V_{eff}}{\partial
g_0^4}\bigg|_{(f_0,g_0)=(0,M)}
\end{equation}
Where M is the renormalised mass. This leads to the condition 
\begin{equation}
\ln\frac{g^2M^2}{\mu^2\sqrt2}+\frac{25}{6}=0
\end{equation}
This can be used to eliminate the dimensionless coupling constant, $g$, from the theory at
the expense of introducing the new dimensionfull variable $M$. We have then traded a
dimensionless parameter
for a dimensionfull, so that we have dimensional transmutation
\cite{Coleman-Weinberg}.
Using the above relation to remove the cutoff, $\mu$, we can write
\begin{eqnarray}
V_{eff}&=&\frac{g^2}{8}(f_0^2-
g_0^2)^2+\frac{22}{3}\frac{1}{16\pi^2}\frac{g^4}{8}(f_0^2-
g_0^2)^2\left(\ln\frac{1}{M^2}|f_0^2-g_0^2|-\frac{25}{6}\right)
\end{eqnarray}
This is our main result.
For consistency, we now verify that this leads to the familiar $\beta$-function for
Yang-Mills. 
If we choose a different  $M$, say $M'$, the coupling constant is defined as:
\begin{equation}
g'^2=\frac{1}{3}\frac{\partial^4V_{eff}}{\partial f_0^4}\bigg|_{(f_0,g_0)=(M',0)}=\frac{1}{3}\frac{\partial^4V_{eff}}{\partial g_0^4}\bigg|_{(f_0,g_0)=(0,M')}
\end{equation}
These two relations, (27) and (30), give
\begin{equation}
g'=\frac{g}{\sqrt{1+g^2\frac{22}{3}\frac{1}{16\pi^2}\ln\frac{M'^2}{M^2}}}
\end{equation}
and the $\beta$-function is
\begin{equation}
\beta=-\frac{22}{3}\frac{g^3}{16\pi^2}
\end{equation}
Hence we conclude that the effective theory in the new variables lead to the correct $\beta$-function for SU(2) Yang-Mills as expected.

We now employ the effective potential (29) to locate the ground state of the theory. 
The minima of the potential are the parabola
\begin{equation}
|f_0^2-g_0^2|=M^2\exp\left(-\frac{24\pi^2}{11g^2}+\frac{11}{3}\right)
\end{equation}
and at the minima the value of the effective potential is
\begin{equation}
V_{eff}(\mbox{min})=-\frac{11}{24}\frac{g^4M^4}{16\pi^2}\exp\left(-\frac{48\pi^2}{11g^2}+\frac{22}{3}\right)
\end{equation} 
In particular we conclude that we have four symmetric non-trivial parabola of
minima, see figure 2. Clearly the origin is not a minima of the
potential any more. This give rise to a symmetry breaking of the
theory and is consistent with the existence of a mass gap in Yang-Mills
theory. Notice that the vacuum remains degenerate at this level of
approximation. Higher order corrections to the effective potential might
result in additional symmetry breaking but that remains to be
investigated.

\vspace{0.3cm}
We have employed the effective potential formalism to Yang-Mills theory
using the Faddeev-Niemi decomposition of the gauge fields. We found that
this leads to a new way of viewing the ground state of the theory in
terms of a potential of two scalar fields. Furthermore we demonstrate
that at the one-loop level of approximation symmetry breaking and
dimensional transmutation occurs. This is consistent with the existence
of a mass gap in Yang-Mills.
It remains to investigate the effect of higher order corrections to the
effective potential. Also it would be useful to verify our results by
explicit computation i.e. without using the arguments of gauge invariance.

\begin{figure}[h]
\epsfig{file=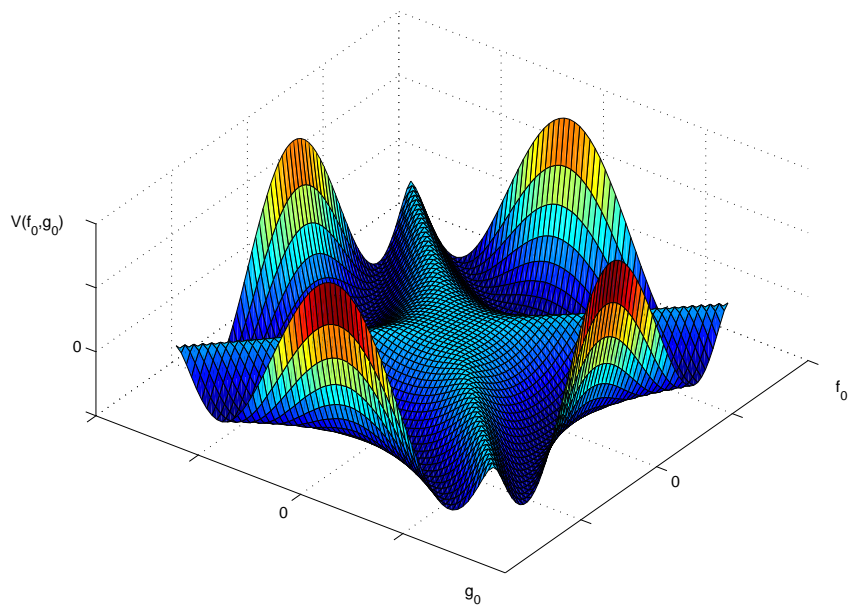, height=8cm}
\caption{The effective potential in terms of the two scalar fields in the Faddeev-Niemi decomposition}
\end{figure}
\vspace{0.3cm}
\noindent
I would like to thank L. Faddeev, E. Langmann and A. Niemi for many valuable discussions on this subject.

\end{document}